\documentclass[preprintnumbers,nofootinbib,a4paper,10pt]{revtex4-1}

\usepackage[utf8x]{inputenc}

\pdfoutput=1

\usepackage{amsmath,latexsym,amssymb,amsfonts}
\usepackage{float} \usepackage{graphicx}
\usepackage{epstopdf}
\usepackage{graphics}
\usepackage{caption}
\usepackage{subfigure}
\usepackage{bm}
\usepackage{color}

\usepackage[dvips]{psfrag}
\usepackage{ulem}
\usepackage{epsfig}
\usepackage{cases}
\usepackage{tensor}
\usepackage{multirow,array}
\usepackage{booktabs} 

\usepackage{colortbl} 

\newcommand{\cro}{\text{cr}}

\newcommand{\DM}{\text{DM}}
\newcommand{\m}{\text{m}}
\newcommand{\TA}{\text{A}}

\newcommand{\tc}{\tilde{c}}

\newcommand{\tG}{\tilde{G}}

\newcommand{\TM}{\text{M}}
\newcommand{\TC}{\text{C}}

\newcommand{\GR}{\text{GR}}

\newcommand{\Tth}{\text{th}}

\newcommand{\Tgr}{\text{gr}}
\newcommand{\Tdyn}{\text{dyn}}

\newcommand{\Tap}{\text{ap}}

\newcommand{\TE}{\text{E}}

\newcommand{\Teff}{\text{eff}}

\newcommand{\Tobs}{\text{obs}}

\def\lambdabar{\ThisStyle{\ensurestackMath{\stackon[-2.4\LMpt]{%
  \SavedStyle\lambda}{\kern-.5pt\kern\LMpt\rule{1\LMex}{.25pt+.15\LMpt}}}}}

\begin{document}

\title{Constraints on the time variation of the speed of light using Strong lensing}
\author{Seokcheon Lee}
\email[Email:]{skylee2@gmail.com}
\affiliation{Department of Physics, Institute of Basic Science, Sungkyunkwan University, Suwon 16419, Korea}

\begin{abstract}
Due to the recent growth of discoveries of strong gravitational lensing (SGL) systems, one can statistically study both lens properties and cosmological parameters from 161 galactic scale SGL systems. We analyze meVSL model with the velocity dispersion of lenses by adopting the redshift and surface mass density depending power-law mass model. Analysis shows that meVSL models with various dark energy models including $\Lambda$CDM, $\omega$CDM, and CPL provide the negative values of meVSL parameter, $b$ when we put the prior to the $\Omega_{\m0}$ value from Planck. These indicate the faster speed of light and the stronger gravitational force in the past. However, if we adopt the WMAP prior on $\Omega_{\m0}$, then we obtain the null results on $b$ within 1-$\sigma$ CL for the different dark energy models.   
\end{abstract}

\maketitle


\section{Introduction}
\label{sec:intro}

Gravitational lensing represents the deflection of light rays owing to intermediate inhomogeneous distributions of matter between the source and the observer. Most sources seem slightly distorted compared to the way they would appear in a perfectly homogeneous and isotropic universe. This is called weak gravitational lensing (WGL) \cite{Refregier:2003ct,Hoekstra:2008db,Bartelmann:2010fz}.

On the other hand, strong gravitational lensing (SGL) implies that the deflection angle is sufficiently large to produce multiple images, arcs, or Einstein rings of a distant background source. SGL systems require massive foreground objects that act as lenses and a background source to be almost perfectly aligned along the line of sight (LOS) from the observer. Any lens with a surface mass density above the so-called critical density can generate multiple images. Hence, the SGL of stars is quite rare. However, SGL is more commonly observed among galaxies and clusters of galaxies. They are massive enough to split multiple images by more than an arcsecond to be resolved by astronomical observations in various wavelengths \cite{Treu:2010}.   

SGL can be used as a powerful method to probe the properties of various astrophysical and cosmological objects. 

\begin{itemize}
	\item \textbf{Mass distribution :} SGL provides an accurate measurement of the total mass (including dark matter, DM) of galaxies and galaxy clusters at cosmological scales. After accounting for the luminous matter in these systems through observations, we can infer the fraction of dark matter and put constraints on its mass distribution.
		\item \textbf{Statistical probes :}
Statistical investigation of lenses their abundances, image separations, and the orientation of their lensed arcs, are useful probes of the foreground lens population. These can be used to constrain physical properties such as shapes of their mass distribution and ellipticities. They can be used to constrain parameters of cosmological models of the Universe.
	\item \textbf{Cosmology :}
Intrinsically variable lensed sources (such as supernovae and quasars) exhibit variabilities among the multiple images at different times due to the different paths taken by light rays to the observer. These time delays can be combined with a lens mass model and lens stellar kinematics to measure both the time-delay distance and the angular diameter distance to the lens, which can constrain cosmological parameters. One can investigate the equation of state (e.o.s) of dark energy (DE) and/or its cosmological evolution by measuring the angular diameter distances of background sources at multiple redshifts behind a lens galaxy or cluster.
\end{itemize}

SGL by elliptical galaxies has been used to probe both cosmology and galaxies \cite{Chae:2002uf,Oguri:2006qp,Paraficz:2009,Paraficz:2010,Cao:2012,Oquri:2012,Chen:2018jcf,Tu:2019vcj,Wong:2019kwg,Oguri:2019fix}. We summarize three main approaches to these methods. 

\begin{itemize}
	\item {\bf Image separation :} SGL systems have been used as an alternative way to probe the property of DE over the years by using measurements for the velocity dispersion of the lens DM halo $\sigma_{\DM}$ and the angular Einstein radius $\theta_{\TE}$ to constrain cosmological parameters \cite{Biesiada:2006,Jullo:2010,Biesiada:2010,Cao:2012,Cao:2015qja,Magana:2015wra,Magana:2017gfs,Amante:2019xao}. It has been demonstrated that the lensing mass distribution of early-type galaxies (ETGs) is close to a singular isothermal sphere (SIS) and thus the separation among the multiple images depends on the mass of the deflector and the angular diameter distances to the lens and to the source when a galaxy acts as a lens. The comoving distance $D_{\TC}$, the transverse comoving distance $D_{\TM}$, and the angular diameter distance $D_{\TA}$ in meVSL model are given by \cite{Lee:2020zts}
\begin{align}
 D_{\text{C}}(z) &\equiv \int_{0}^{r} \frac{dr'}{\sqrt{1-kr^2}} = \frac{\tc_0}{H_0} \int_{0}^{z} \frac{dz'}{E^{(\GR)}(z')} \quad \text{where} \quad  E^{2} \equiv \left( \sum_{i} \Omega_{i0} a^{-3(1+\omega_i)} \right)  \label{Dcmp} \,, \\
 D_{\text{M}}(z) &=  \begin{cases} \frac{\tc_0}{H_0} \frac{1}{\sqrt{\Omega_{K0}}} \sinh \left( \sqrt{\Omega_{k0}} \frac{H_0}{\tc_0} D_{\text{C}} \right) & \Omega_{k0} > 0 \, \\D_{\text{C}} & \Omega_{k0} = 0 \, \\ \frac{\tc_0}{H_0} \frac{1}{\sqrt{|\Omega_{K0}|}} \sin \left( \sqrt{|\Omega_{k0}|} \frac{H_0}{\tc_0} D_{\text{C}} \right) & \Omega_{k0} < 0 \, \end{cases} \label{DMmp} \,, \quad D_{\TA}(z) = \frac{1}{(1+z)} D_{\TM}(z) 
 \end{align}	
 where $H_0$ is the present value of the Hubble parameter, $\tc_0$ is the present value of the speed of light, $\Omega_{i0} \equiv \rho_{i 0} / \rho_{\cro 0}$ is the present mass density contrasts of i-component, and $\omega_i$ is its equation of state (e.o.s).
	
One defines $\mathcal R^{\Tth}$ as the theoretical angular diameter distance ratio of the angular diameter distance between the lens and source $D_{\TA}(z_l,z_s)$ and that between observer and source $D_{\TA}(0,z_s)$. Also, its observable counterpart $\mathcal R^{\Tobs}$ can be defined as
		\begin{align}
		\mathcal R^{\Tth} \left( z_{l} , z_{s} ; {\bf p} \right) &= \frac{D_{\TA} \left( z_l, z_s ; {\bf p} \right)}{D_{\TA} \left( 0, z_s ; {\bf p} \right)} = \frac{D_{\TM} \left( z_l, z_s ; {\bf p} \right)}{D_{\TM} \left( 0, z_s ; {\bf p} \right)}  \label{Rth} \,, \\
		 \mathcal R^{\Tobs} \left( \sigma_{\DM} , \theta_{\TE} \right) &= \frac{\tc_0^2 \theta_{\TE}}{4 \pi \sigma_{\DM}^2} \label{Robs} \,. 
		\end{align}
where $\textbf{p} = \sum_{i} (\Omega_{i0} , \omega_{i})$ denote cosmological parameters, $z_{l}$ and $z_{s}$ are the redshifts to the lens and source respectively. Thus, one can estimate cosmological parameters by minimizing the $\chi^2$ function by using observed values of $z_{l}$, $z_s$, $\theta_{\DM}$, and $\theta_{\TE}$ from SGL systems
	\begin{align}
	\chi^2 (\textbf{p}) = \sum_{i=1}^{N_{\text{SGL}}} \left( \frac{\mathcal R_{i}^{\Tth} \left( z_{l} , z_{s} ; {\bf p} \right) - \mathcal R_{i}^{\Tobs}\left( \sigma_{\DM} , \theta_{\TE} \right)}{ \Delta \mathcal R_{i}^{\Tobs}} \right)^2 \quad , \quad \Delta \mathcal R_{i}^{\Tobs} \equiv \mathcal R_{i}^{\Tobs} \sqrt{ \left( \frac{\Delta \theta_{\TE}}{\theta_{\TE}} \right)_i^2 + \left( 2 \frac{\Delta \sigma_{\DM}}{\sigma_{\DM}}  \right)_i^2 } \label{chi2Rth} \,,
\end{align}
 where $\Delta \theta_{\TE}$ and $\Delta \sigma_{\DM}$ are the errors for the angular Einstein distance and for the DM velocity dispersion respectively. 
 
	\item {\bf Time delay cosmography :} In the SGL systems, multiple images appear to the observer. If the emission intensity from the source is variable in time as an active galactic nucleus (AGN), a supernova (SN), or a quasar, the difference in arrival time is measurable. This arrival time of the images depends on both the path of the light ray and also the gravitational potential of the lens \cite{Suyu:2012ApJ}. The light travel time $t \left( \boldsymbol{\theta} \,, \boldsymbol{\beta} \right) $ for each image at angular position $\boldsymbol{\theta}$ relative to a fiducial unperturbed ray from the source position $\boldsymbol{\beta}$ through a single, isolated, thin gravitational lens is given by 
	\begin{align}
	t \left( \boldsymbol{\theta} \,, \boldsymbol{\beta} \right) &\equiv \frac{D_{\Delta t}}{c} \phi \left( \boldsymbol{\theta} \,, \boldsymbol{\beta} \right)  \label{timet}  \,, 
	\end{align}
	where a so-called ``time-delay distance'' $D_{\Delta t}$ and the Fermat potential $\phi \left( \boldsymbol{\theta} \,, \boldsymbol{\beta} \right) $ are defined as
	\begin{align}
	D_{\Delta t} &\equiv \frac{D_{\TA}(0,z_l) D_{\TA}(0,z_s)}{D_{\TA}(z_l,z_s)} \label{DDeltat} \,, \\
	\phi \left( \boldsymbol{\theta} \,, \boldsymbol{\beta} \right) &\equiv \left[ \frac{ \left( \boldsymbol{\theta} - \boldsymbol{\beta}  \right)^2 }{2} - \psi \left( \boldsymbol{\theta} \right) \right] \label{FermatPotential} \,.
	\end{align}
	The deflection angle of light rays is determined by the lens equation $\boldsymbol{\theta} - \boldsymbol{\beta} \equiv \boldsymbol{\alpha}$ which is related to the lens potential as $\boldsymbol{\alpha} = \boldsymbol{\nabla} \psi$. The first term of Eq.~\eqref{FermatPotential} comes from the geometric path difference as a result of the strong lens deflection and the second term is due to the gravitational delay described by the lens potential $\psi$. Therefore, one can estimate cosmological parameters including the present value of the Hubble parameters that appear in the time delay distance by modeling the lens potential and the source position. There have been quasar-galaxy strong lensing systems with time delay observations \cite{Suyu:2013kha,Chen:2016fwu,Wong:2016dpo,Bonvin:2018dcc,Birrer:2018vtm,Bonvin:2019xvn,Rusu:2019xrq,Chen:2019ejq,Wong:2019kwg}. $H_0$ Lenses in COSmological MOnitoring of GRAvItational Lenses (COSMOGRAIL)'s Wellspring (H0LiCOW) collaboration reveals six lensed quasars and one from STRong lensing Insights into the Dark Energy Survey (STRIDES) collaboration \cite{Shajib:2019toy}. The eight lenses with measured time delays consist of the following systems : B1608+656 \cite{Myers:1995,Fassnacht:1996,Suyu:2009by,Jee:2019hah}, RXJ0911+0551 \cite{Bade:1997,Kneib:2000ty,Hjorth:2002gs}, RXJ1131-1231 \cite{Sluse:2003iy,Sluse:2007cn}, HE0435-1223 \cite{Wisotzki:2002dp,Morgan:2004xu,Eigenbrod:2005ub,Sluse:2012rg,Millon:2020xab}, SDSS1206+4332 \cite{Oguri:2004qu,Agnello:2015ala}, WFI2033-4723 \cite{Sluse:2012rg,Morgan:2003bd,Sluse:2019}, PG1115+080 \cite{Weymann:1980,Tonry:1997pc}, and DES J0408-5354 \cite{Shajib:2019toy}. Recently, there are also time delays in 18 strongly lensed quasars from optical monitoring \cite{Millon:2020xab}.
	
	\item {\bf Lens velocity dispersion :} If we combine SGL observations with stellar dynamics in elliptical galaxies, then we can use the spectroscopically measured stellar velocity dispersion (VD) of the lens as statistical quantity to constrain both cosmological parameters and lens model \cite{Futamase:2001ij,Biesiada:2006,Grillo:2007iv,Schwab:2010,Cao:2017nnq,Chen:2018jcf,Amante:2019xao}. The main assumption is that the dynamical mass enclosed within the Einstein radius and lensing mass of early-type lens galaxies should be equal. Available SGL systems of VD have grown \cite{Koopmans:2002ic,Treu:2002ee,Koopmans:2002qh,Treu:2004wt,Bolton:2005nf,Treu:2005aw,Koopmans:2006iu,Grillo:2007iv,Ruff:2011,Brownstein:2012,Gavazzi:2012,Sonnenfeld:2013xga,Sonnenfeld:2013cha,Sonnenfeld:2014,Cao:2015qja,Zhou:2019vou}. We use this method to analyze the cosmological parameters of meVSL model in this manuscript. Thus, we will briefly review the methodology of this approach in the next section. \end{itemize} 

The outline of this manuscript is as follows. In Section \ref{sec:Rev}, we briefly review the method of using stellar VD as statistical quantity and derive the modifications of equations in meVSL model. We analyze data to obtain cosmological parameters and time evolutions of the speed of light and the gravitational constant in Section \ref{sec:Anal}. In Section \ref{sec:Con}, we conclude and summarize. 

\section{Review on VD method}
\label{sec:Rev}

\begin{figure}
\centering
\includegraphics[width=0.9\linewidth]{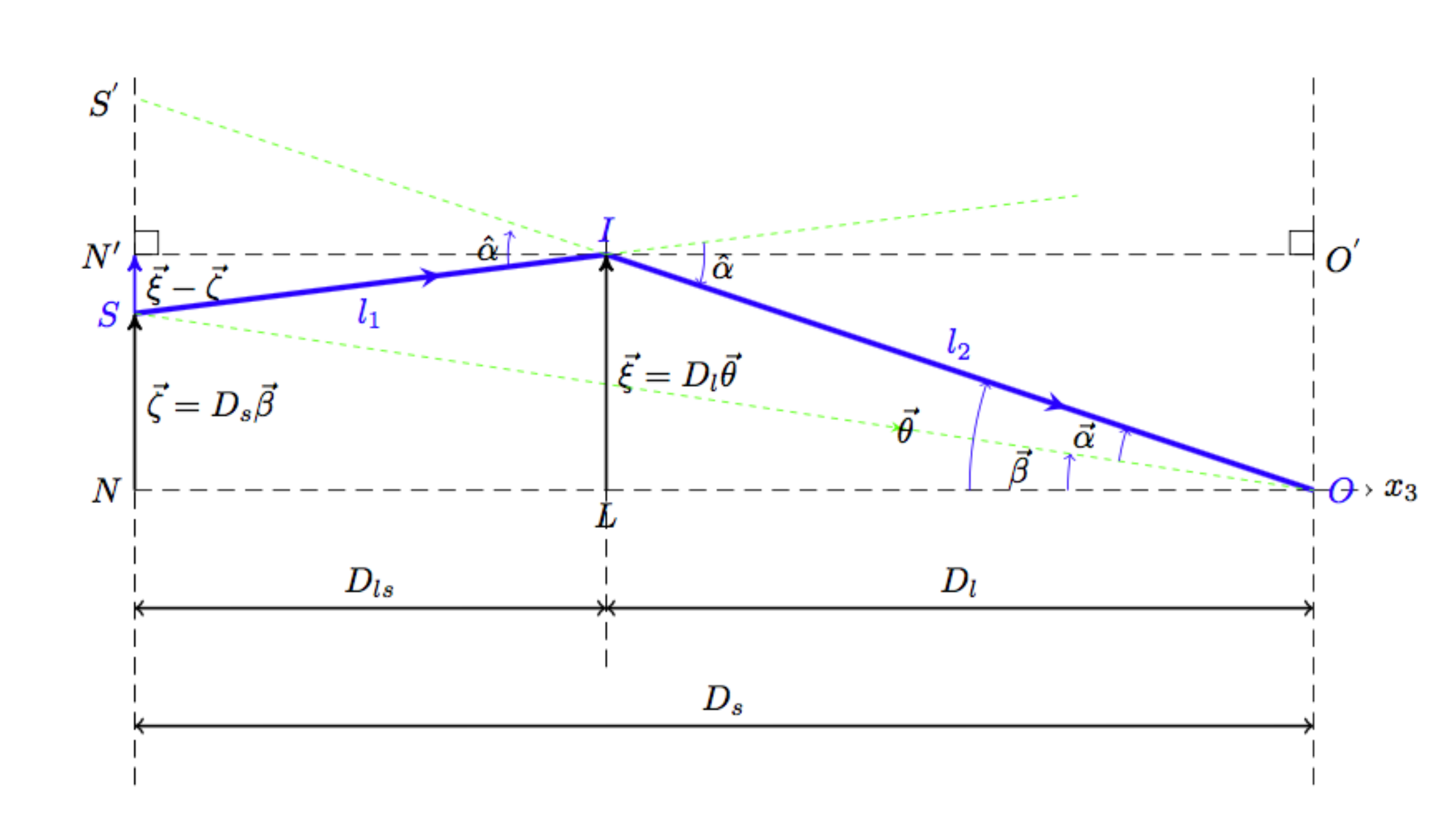} 
\vspace{-0.5cm}
\caption{Configuration of strong gravitational lensing system.}
\label{Fig:1}
\end{figure}

A typical configuration considered in SGL is depicted in Fig.~\ref{Fig:1}, where a lens at distance $D_{l}$ deflects the light rays from a source at distance $D_{s}$. The source and lens planes are perpendicular to the optical axis (line $NO$) from the source to the observer. $\vec{\zeta}$ denotes the two-dimensional position of the source on the source plane and it is related to the true angular position $\vec{\beta} = \vec{\zeta} / D_{s} $ in the absence of the deflector. If there are no other lenses close to the LOS and the extent of the lens mass along it is much smaller than both $D_l$ and the distance from the deflector to the source $D_{ls}$, the actual light rays which are smoothly curved in the vicinity of the deflector can be replaced by two straight rays with the deflection angle $\hat{\alpha}$. This is also related with the scaled deflection angle $\vec{\alpha} = (D_{ls}/D_{s}) \hat{\alpha}$. The lens equation relates the true position of the source to its observed position, $\vec{\beta} = \vec{\theta} - \vec{\alpha}$ as shown in Fig.~\ref{Fig:1}. 

Among several methods of using SLG to probe cosmology, the method of using stellar VD has merit since its theoretical prediction depends only on the lens mass model without relying on the luminosity function. We adopt the method in \cite{Chen:2018jcf} and thus briefly review it in this section and also extend results of GR to those of the meVSL model.

The main assumption of this method is that the projected surface mass due to the lens $M^{\TE}_{\Tgr}$ equals the projected dynamical mass $M^{\TE}_{\Tdyn}$ with the Einstein radius $R_{\TE} = D_{l} \theta_{\TE}$. The projected gravitational mass is related with the critical surface mass density by $\pi \Sigma_{\cro} R_{\TE}^2 = M^{\TE}_{\Tgr}$ where $\Sigma_{\cro} = c^2/(4 \pi G) \cdot D_{s}/(D_{s} D_{ls})$. Thus, the projected gravitational surface mass of meVSL model is generalized as
\begin{align}
M^{\TE}_{\Tgr} = \frac{\tc_0^2}{4 \pi \tG_{0}}  (1 + z_l)^{-\frac{b}{2}} \frac{D_{s} D_{l}}{D_{ls}} \theta_{\TE}^2 \label{Mgr} \,,
 \end{align}  
where $\tc_0$ and $\tG_0$ correspond to the present values of the speed of light and of the gravitational constant, respectively. We use the consequences of meVSL model $\tc = \tc_0 (1+z)^{-b/4}$ and $\tG = \tG_0 (1+z)^{b}$. One should make an assumption on the galaxy lens mass distribution in order to estimate the projected dynamical mass of the lens galaxy from its VD. The density distribution of total mass $\rho(r)$ assumed to follow that of early-type galaxies (ETGs) and the luminosity density of stars $\nu(r)$ is given by
\begin{align}
\rho(r) &= \rho_0 \left( \frac{r}{r_0} \right)^{-\gamma(z)} \quad \text{where} \quad \gamma = \gamma_0 + \gamma_{z} \cdot z_{l} + \gamma_s \cdot \log \tilde{\Sigma} \label{rhor} \,, \\
\nu(r) &= \nu_0 \left( \frac{r}{r_0} \right)^{\delta} \quad , \quad 
\beta(r) = 1 - \frac{\sigma_{\theta}^2}{\sigma_{r}^2} \label{betar} \,, 
\end{align} 
where $\tilde{\Sigma}$ is the normalized surface mass density of the lens galaxy given by $(\sigma/100 \, \text{km} \, \text{s}^{-1})^2/(R_{\TE}/10 h^{-1} \, \text{kpc})$ and $\beta(r)$ denotes the stellar orbital anisotropy with $\sigma_{\theta}$ and $\sigma_{r}$ represent the tangential and radial VDs, respectively.  

The projected dynamical mass contained within a cylinder with a radius equals to $R_{\TE}$ with the density distribution given in Eq.~\eqref{rhor} is
\begin{align}
M^{\TE}_{\Tdyn} &= 2 \sqrt{\pi^{3}} \frac{R_{\TE}^{3-\gamma}}{3 - \gamma} \frac{\Gamma \left[ \left( \gamma-1 \right)/2 \right]}{\Gamma \left[ \gamma/2\right]} \rho_{0} r_0^{\gamma} \,. \label{MTEdyn} 
\end{align}
The total mass contained within a sphere of radius $r$ related to $M^{\TE}_{\Tdyn}$ can be written as
\begin{align}
M(r) = 4 \pi \frac{r^{3-\gamma}}{3 - \gamma}  \rho_{0} r_0^{\gamma}  = \frac{2}{\sqrt{\pi}} \frac{\Gamma \left[ \gamma/2\right]} {\Gamma \left[ \left( \gamma-1 \right)/2 \right]} \left( \frac{r}{R_{\TE}} \right)^{3-\gamma} M^{\TE}_{\Tdyn} \label{Mr} 
\end{align}

One can obtain the radial component of VD, $\sigma_{r}$ by using the radial Jeans equation in spherical coordinate under the assumption that the stellar number density is spatially constant compared to stellar luminosity density $\nu(r)$ 
\begin{align}
\sigma_{r}^2 (r) &= \tG_{0} \left( 1 + z_l \right)^{-b} \int_{r}^{\infty} dr' r^{' 2 \left( \beta - 1\right)} \nu \left( r'\right) M(r') / \left( r^{2 \beta} \nu (r) \right) \label{sigmar2} \,.
\end{align}
The final expression of radial VD is obtained by inserting $M(r)$ in Eq.~\eqref{Mr} into Eq.~\eqref{sigmar2} 
\begin{align}
\sigma_{r}^2(r) &= \frac{2}{\sqrt{\pi}} \frac{G M^{\TE}_{\Tdyn}}{R_{\TE}} \frac{1}{\eta - 2\beta} \frac{\Gamma \left[ \gamma/2\right]} {\Gamma \left[ \left( \gamma-1 \right)/2 \right]} \left( \frac{r}{R_{\TE}} \right)^{2-\gamma} \,, \label{sigmar22} 
\end{align}
where $\eta = \gamma + \delta -2$. The actually measured VD of the lens galaxy is the component of luminosity weighted average along the LOS and over the effective spectroscopic aperture $R_{\TA}$ and the final expression for this is given by
\begin{align}
\sigma_{\parallel}^2 \left( \leq R_{\TA} \right) &= \frac{2}{\sqrt{\pi}} \frac{\tG_0 M^{\TE}_{\Tdyn}}{R_{\TE}}  f \left( \gamma \,, \beta \,, \delta \right)  \left( \frac{R_{\TA}}{R_{\TE}} \right)^{2 - \gamma} (1+z_l)^{-b} \nonumber \\
		&= \frac{\tc_0^2}{ 2 \sqrt{\pi}} \frac{D_{s}}{D_{ls}} \theta_{\TE} f \left( \gamma \,, \beta \,, \delta \right)  \left( \frac{\theta_{\Teff}}{2 \theta_{\TE}} \right)^{2 - \gamma} (1+z_l)^{-\frac{b}{2}}  \,, \quad \text{where} \label{sigmap2} \\
f \left( \gamma \,, \beta \,, \delta \right) &\equiv \frac{3 - \delta}{\left( \eta - 2 \beta \right) \left( 3 - \eta \right)} \left[ \frac{\Gamma \left[ \left( \eta-1 \right)/2 \right]}{\Gamma \left[ \eta/2\right]} - \beta \frac{\Gamma \left[ \left( \eta+1 \right)/2 \right]}{\Gamma \left[ \left( \eta+2 \right)/2  \right]} \right] \frac{\Gamma \left[ \gamma/2 \right] \Gamma \left[ \delta/2 \right]}{\Gamma \left[ \left( \gamma -1 \right)/2 \right] \Gamma \left[ \left( \delta -1 \right)/2 \right]} \,, \nonumber 
\end{align}
where we also use $\theta_{\TA} = \theta_{\Teff}/2$. Thus, Eq.~\eqref{sigmap2} is the theoretical value of radial VD. The corresponding observational value of VD is obtained from the spectroscopic data $\sigma_{\Tap}$ inside the circular aperture with the angular radius $\theta_{\Tap}$. From the aperture correction formula, one can estimate the observational radial VD as
\begin{align}
\sigma_{\parallel}^{\Tobs} = \sigma_{\Tap} \left( \frac{\theta_{\Teff}}{2 \theta_{\Tap}} \right)^{\tau} \quad \text{where} \quad \tau = -0.066 \pm 0.035 \label{sigmaobs} \,,
\end{align}
where the value of $\tau$ is from \cite{Cappellari:2005ux}. There are 161 galaxy scale SGL systems with given observed values of $z_s$, $z_l$, $\theta_{\TE}$, $\theta_{\Teff}$, $\theta_{\Tap}$, and $\sigma_{\Tap}$ with its error. Thus, we can perform the likelihood analysis by doing the minimum $\chi^2$ for VD as
\begin{align}
\chi^2 = \sum_{i=1}^{N} \left( \frac{\sigma_{\parallel \,, i}^{\Tth} - \sigma_{\parallel \,, i}^{\Tobs} }{\Delta  \sigma_{\parallel \,, i}^{\Tobs}}  \right)^2 \label{chi2}  \,,
\end{align}  
where $N$ is the number of the data points, $\sigma_{\parallel \,, i}^{\Tth}$ is given in Eq.~\eqref{sigmap2}, $\sigma_{\parallel \,, i}^{\Tobs}$ is the observed value of VD given in Eq.~\eqref{sigmaobs}, and $\Delta  \sigma_{\parallel \,, i}^{\Tobs}$ is its uncertainty.

\section{Analysis and results}
\label{sec:Anal}

We analyze three different DE models for two different given values of $\Omega_{\m0}$ by using the data given in \cite{Chen:2018jcf}. We investigate the cosmological constant ($\omega_0 = -1$), the constant $\omega_0$ (\textit{i.e.} $\omega_a = 0$), and the Chevallier-Polarski-Linder (CPL) parametrization ($\omega = \omega_{0} + \omega_{a} (1-a)$) for the DE equation of state (e.o.s) as viable DE models under the meVSL model. We adopt $\Omega_{\m0} = 0.27$ from the Wilkinson Microwave Anisotropy Probe (WMAP) seven-year results \cite{Jarosik:2010} and $\Omega_{\m0} = 0.32$ from 2018 Planck data release \cite{Aghanim:2018eyx}. The overall results for different models are given in Table.~\ref{tab:omegaw0}. Overall results show that the speed of light is slower than that of present and the gravitational constant is smaller than that of the present ( \textit{i.e.} $b > 0$) when we adopt the WMAP value of the matter density $\Omega_{\m0} = 0.27$. Within 1-$\sigma$ CL those variations are null. However, if we adopt the matter density from the Planck $\Omega_{\m0} = 0.32$, then we obtain the larger values both of the speed of light and of the gravitational constant in the past ( \textit{i.e.} $b > 0$). Usually, there is no dependence on $h$ based on the VD analysis method because it depends on the ratio of angular diameter distances. However, $h$ dependence appears in the $\gamma$ due to its dependence on $\tilde{\Gamma}$ as shown in Eq.~\eqref{rhor}.  

\subsection{Cosmological constant :  $\Lambda$}
\label{subsec:Lambda}

In this subsection, we constrain the meVSL model parameter $b$ based on the cosmological constant, $\omega_0 = -1$ and $\omega_a = 0$ by using the minimum chi-square analysis. We first fix the best fit values of lens model parameters for the different values of $\Omega_{\m 0}$.

	\begin{itemize} 
		\item \textbf{$\Omega_{\m0} = 0.27$ :} We obtain the best-fit values of lens model parameters as $\left( \gamma_0 \,, \gamma_{z} \,, \gamma_{s} \,, \beta \,, \delta \right) = \left( 1.201 \,, -0.568 \,, 0.815 \,, 2.349 \,, -0.022 \right) $ with $h = 0.717$  and $\chi^2 = 344.8$ for $b = 0$. If we consider this model for meVSL model, then we obtain the other best-fit values $\left( \gamma_0 \,, \gamma_{z} \,, \gamma_{s} \,, \beta \,, \delta \right) = \left( 1.182 \,, -0.575 \,, 0.818 \,, 2.370 \,, -0.019 \right) $ with $h = 0.738$. For the 1-$\sigma$ confidence level (CL), $b = 0.032 \pm 0.092$ and $\chi^2 = 359.6$ for this model. 	Thus, we obtain the null results for the time variation of the physical constants like the speed of light and the gravitational constant in this model.  	
		\item \textbf{$\Omega_{\m0} = 0.32$ :} In this model, we obtain $\left( \gamma_0 \,, \gamma_{z} \,, \gamma_{s} \,, \beta \,, \delta \right) = \left( 1.184 \,, -0.589 \,, 0.882 \,, 2.362 \,, -0.018 \right) $ with $h = 0.721$ and $\chi^2 = 431.1$ for $b =0$. If we allow $b$ to vary, then the best fit values become $\left( \gamma_0 \,, \gamma_{z} \,, \gamma_{s} \,, \beta \,, \delta \right) = \left( 1.183 \,, -0.588 \,, 0.822 \,, 2.363 \,, -0.018 \right) $ , $h = 0.731$, and $b = -0.339 \pm 0.091$ with  $\chi^2 = 421.4$. In this model, the lens model parameters are almost independent of the value of $b$. We obtain the cosmological evolutions of the speed of light and that of the gravitational constant in this model. This result is shown in Fig.~\ref{fig-2}. In the left panel of this figure, we depict the cosmological evolution of $\tc/\tc_0$ as a function of the redshift $z$. The central dashed line indicates the best fit value of $\tc/\tc_0$. Its value increases as a function of $z$. The shaded region enveloped by the solid lines represents the 1-$\sigma$ CL region. The cosmological evolution of $\tG/\tG_0$ within 1-$\sigma$ CL is shown in the right panel of Fig.~\ref{fig-2}.   	
\begin{figure*}
\centering
\vspace{1cm}
\begin{tabular}{cc}
\includegraphics[width=0.48\linewidth]{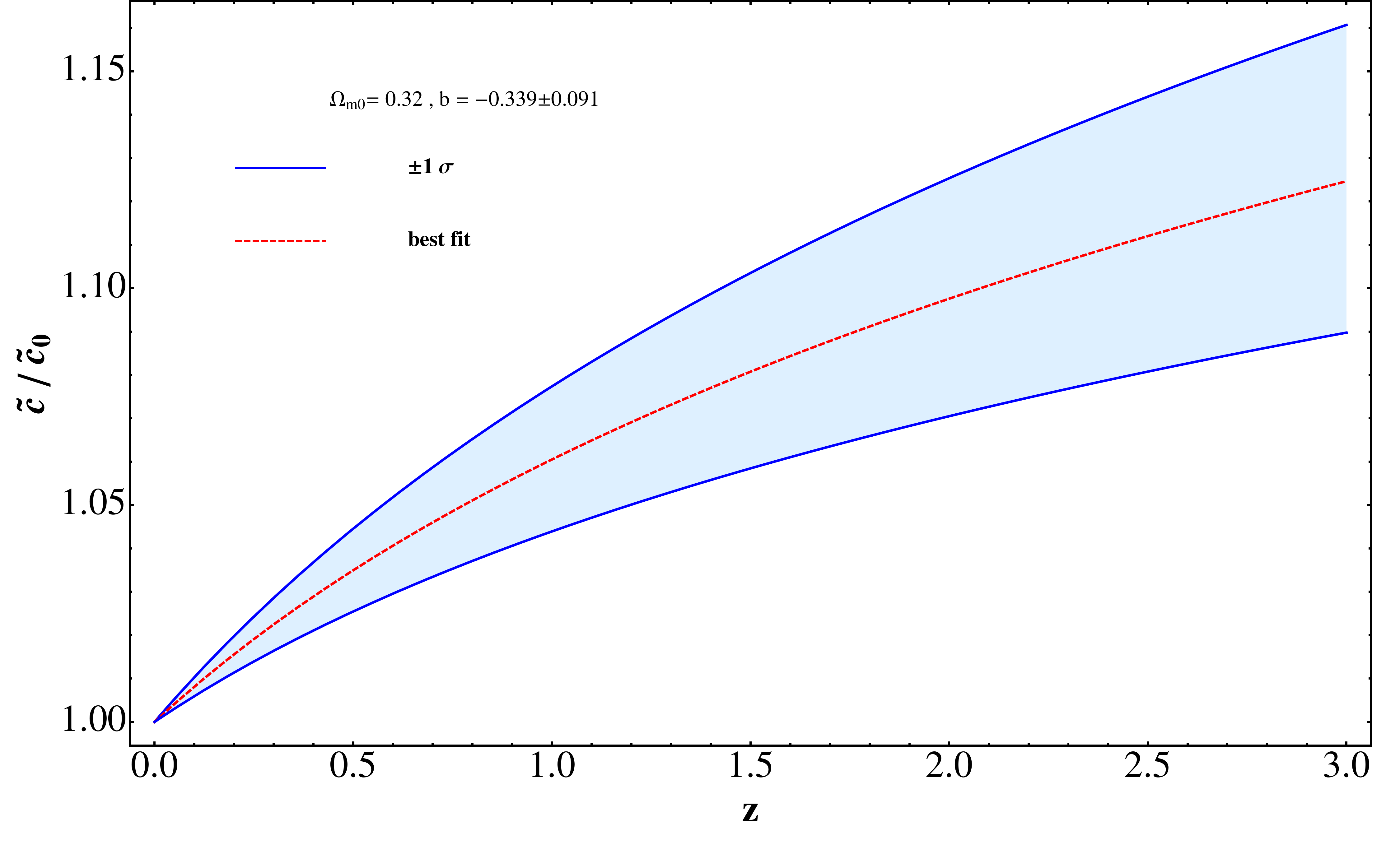} &
\includegraphics[width=0.48\linewidth]{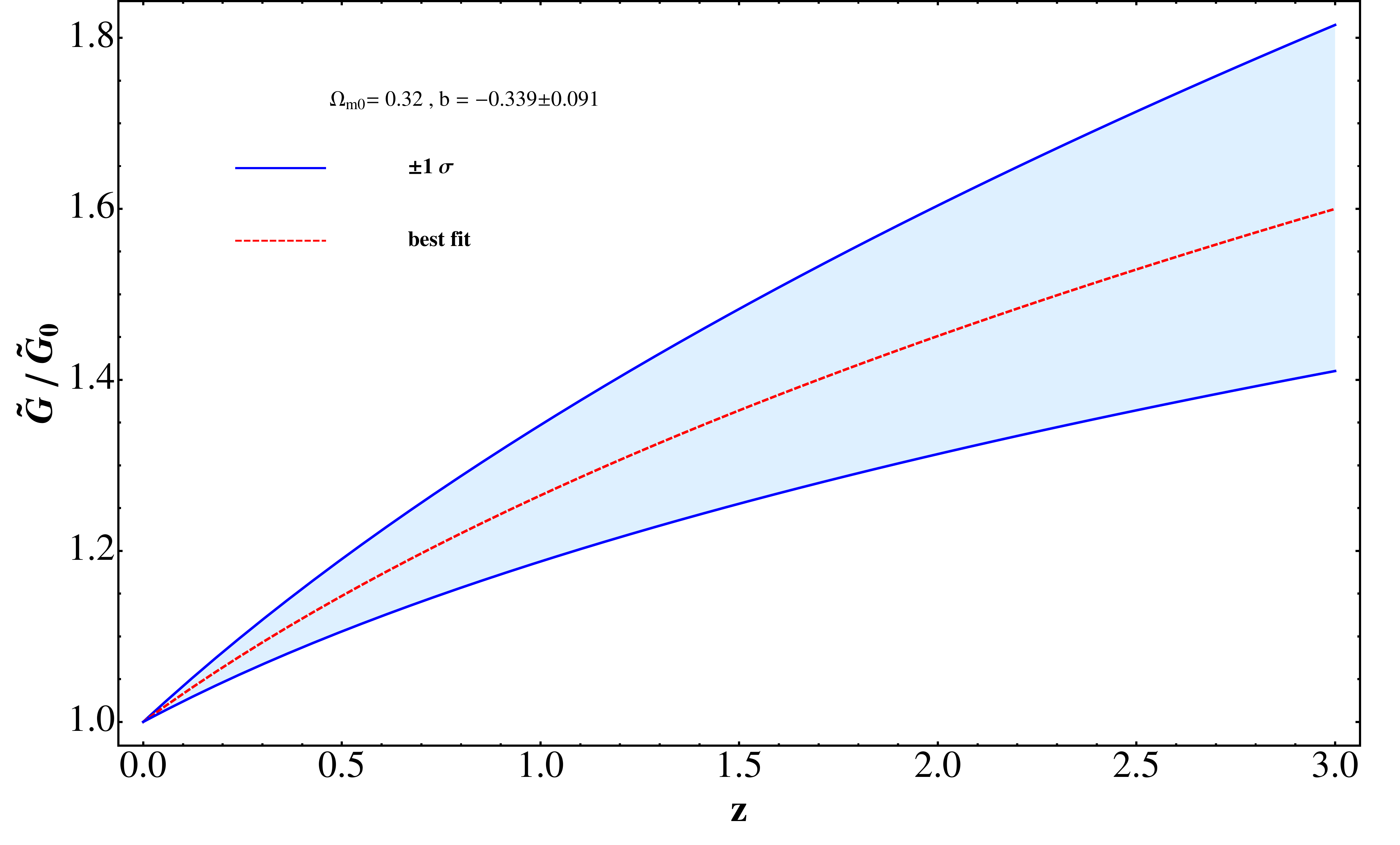} 
\end{tabular}
\vspace{-0.5cm}
\caption{The time variations of $\tc$ and $\tG$ of $\Lambda$CDM meVSL model as a function of $z$ when $\Omega_{\m0} = 0.32$. a) The cosmological evolution of the ratio of $\tc$ to its present value $\tc_0$ within 1-$\sigma$ CL. b) The cosmological evolution of the value of $\tG/\tG_0$ for 1-$\sigma$ CL.} \label{fig-2}
\vspace{1cm}
\end{figure*}
		\end{itemize}

\subsection{Constant DE e.o.s : $\omega$}
\label{subsec:omega}

We investigate the so-called $\omega$CDM model ({\it i.e.} constant $\omega_{0}$ and $\omega_{a} = 0$) in this subsection. Again, we obtain the best fit values of lens model parameters first and then do the likelihood analysis for the $\omega_{0}$ and $b$ values for the different values of $\Omega_{\m0}$.

	\begin{itemize} 
		\item \textbf{$\Omega_{\m0} = 0.27$ :} For $b =0$, the best fit values of $\left( \gamma_0 \,, \gamma_{z} \,, \gamma_{s} \,, \beta \,, \delta \right) = \left( 1.197 \,, -0.570 \,, 0.816 \,, 2.352 \,, -0.021 \right)$ with $h = 0.717$. The 1-$\sigma$ CL of $\omega_0$ is $-0.953 \pm 0.044$ with the minimum chi-square value is $345.8$. We also obtain the other set of the lens model parameters best fit values $\left( \gamma_0 \,, \gamma_{z} \,, \gamma_{s} \,, \beta \,, \delta \right) = \left( 1.198 \,, -0.574 \,, 0.817 \,, 2.367 \,, -0.006 \right) $. The value of $h = 0.710$ with $\chi^2 = 357.0$. Within the 1-$\sigma$ CL, $\omega_0 = -1.006 \pm 0.062$ and $b = 0.127 \pm 0.120$. Interestingly, $\omega_0$ is the same as the $\Lambda$CDM model but this model provides the time-varying speed of light within 1-$\sigma$ CL. We also investigate the 1-$\sigma$ confidence level values of $\omega_0$ and $b$ when we adopt the same best fit values of lens model parameters $\left( \gamma_0 \,, \gamma_{z} \,, \gamma_{s} \,, \beta \,, \delta \right) = \left( 1.18 \,, -0.49 \,, 0.81 \,, 2.48 \,, 0 \right)$ for the different values of $\Omega_{\m0} = 0.32$. In this model, we obtain $(\omega_0 \,, h \,, b ) = (-0.944 \pm 0.054 \,, 0.743 \pm 0.009 \,, 0.331 \pm 0.171)$ with $\chi^2 = 557.4$.  
		\item \textbf{$\Omega_{\m0} = 0.32$ :} The best fit values of $\left( \gamma_0 \,, \gamma_{z} \,, \gamma_{s} \,, \beta \,, \delta \right) = \left( 1.181 \,, -0.591 \,, 0.822 \,, 2.364 \,, -0.017 \right)$ with $h = 0.721$ for $b = 0$. The 1-$\sigma$ CL of $\omega_0$ is $-0.992 \pm 0.042$ with $\chi^2 = 431.1$. For $b \neq 0$, we obtain the best fit values of the lens model parameters as $\left( \gamma_0 \,, \gamma_{z} \,, \gamma_{s} \,, \beta \,, \delta \right) = \left( 1.205 \,, -0.590 \,, 0.821 \,, 2.359 \,, -0.017 \right) $. The best value of $h$ is $0.700$ and $\chi^2 = 418.1$. Within the 1-$\sigma$ CL, $\omega_0 = -0.996 \pm 0.052$ and $b = -0.246 \pm 0.108$. The 1-$\sigma$ confidence level values of $\omega_0$, $h$, and $b$ for the same best fit values of lens model parameters of the different values of $\Omega_{\m0} = 0.27$ become $(-0.914 \pm 0.046 \,, 0.749 \pm 0.009 \,, -0.214 \pm 0.166)$ with $\chi^2 = 649.7$.  We show this result in Fig.~\ref{fig-3}. In the left panel of this figure, we show the constraint contours for $b$-$\omega_0$. The best fit value of $(b \,, \omega_0) = (-0.214 \,, -0.914)$ and the different color shaded regions represent the 1-$\sigma$, 2-$\sigma$, and 3-$\sigma$ constraints. These two parameters show a stronger negative correlation than those with $\omega_0$. We depict the contours of $b$-$h$ constraints for the different confidence levels in the middle panel of Fig.~\ref{fig-3}. $h$-$\omega_0$ constraint contours are shown in the right panel of this figure. 
		
\begin{figure*}
\centering
\vspace{1cm}
\begin{tabular}{ccc}
\includegraphics[width=0.3\linewidth]{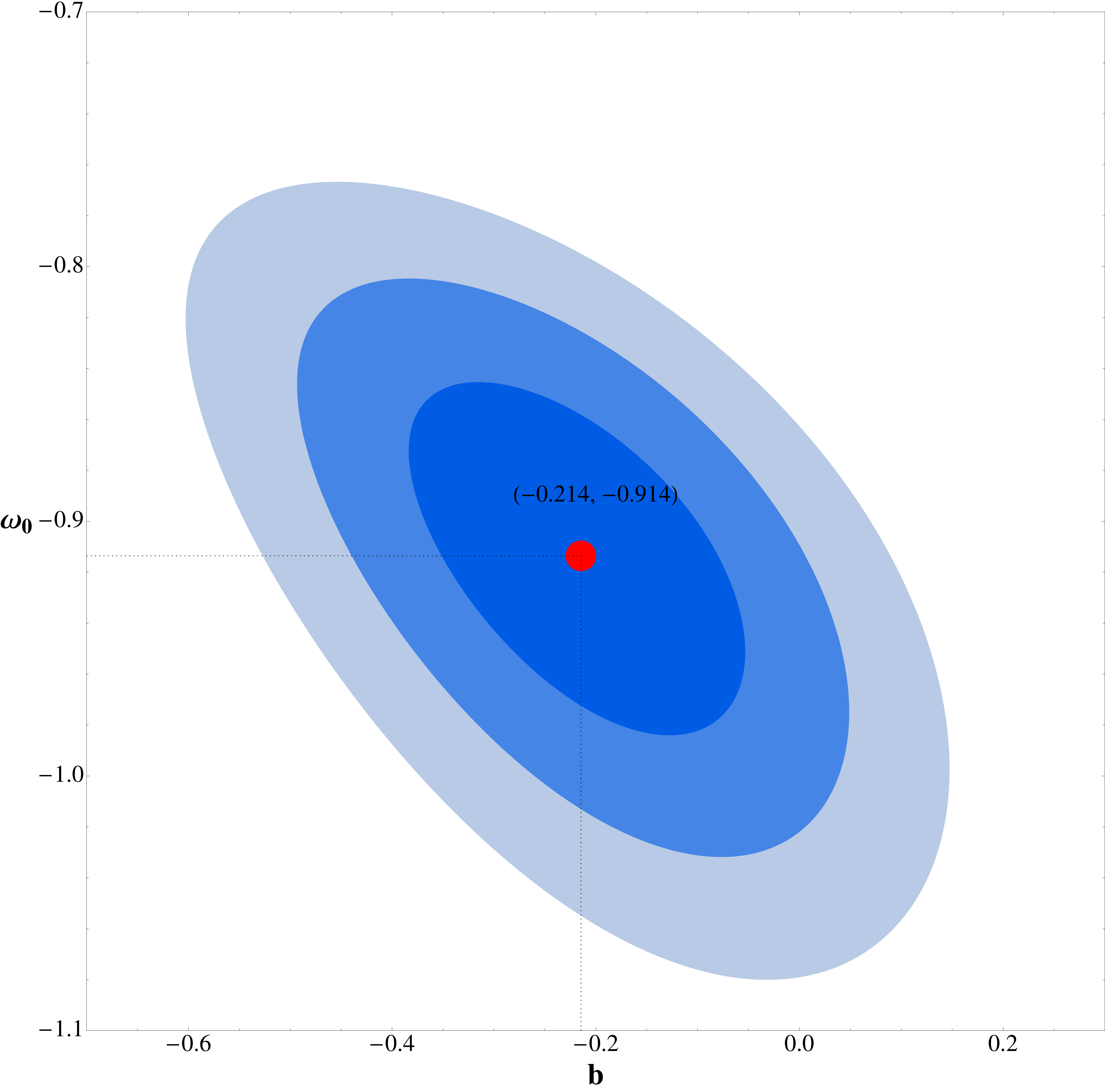} &
\includegraphics[width=0.3\linewidth]{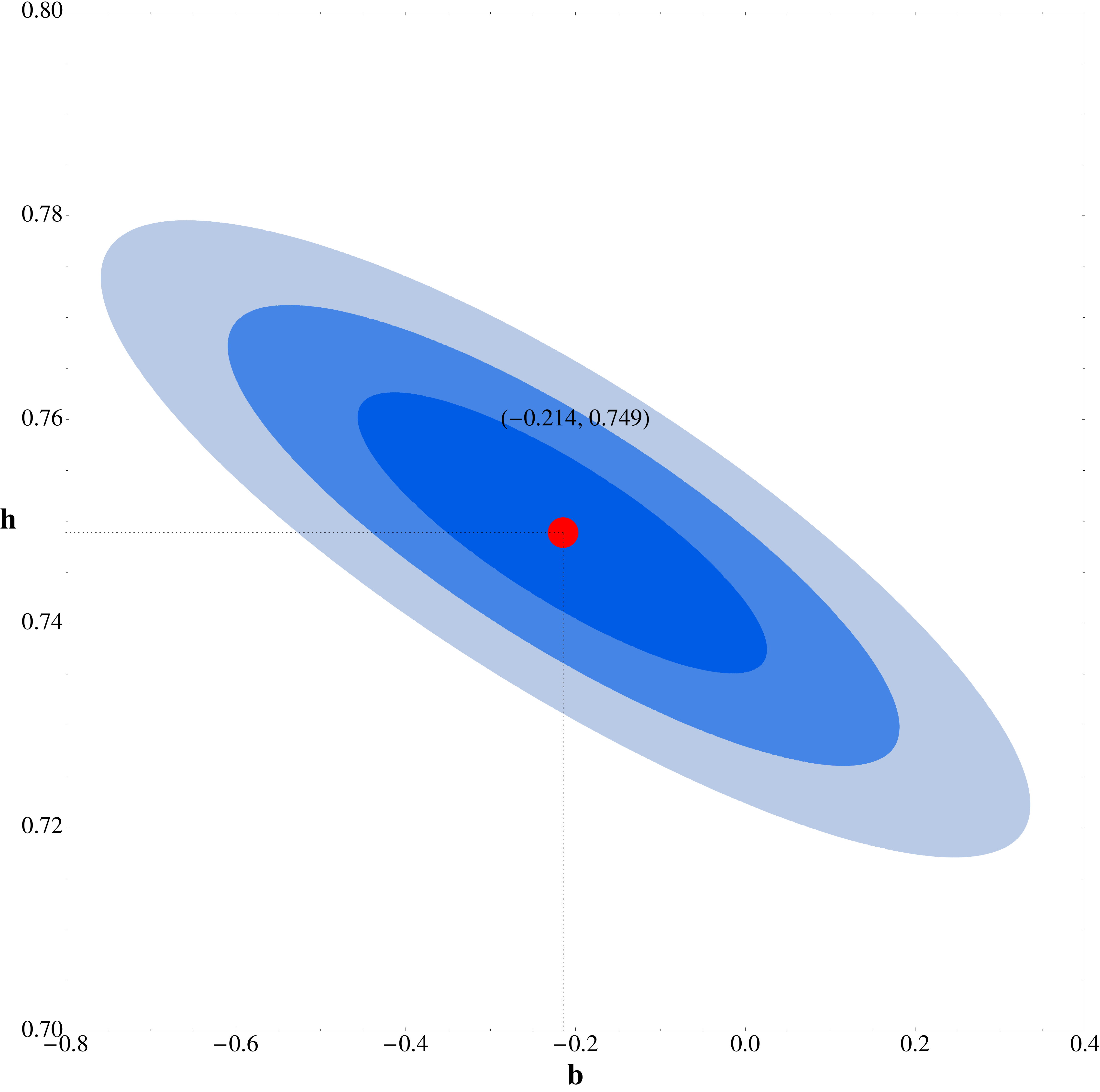} &
\includegraphics[width=0.3\linewidth]{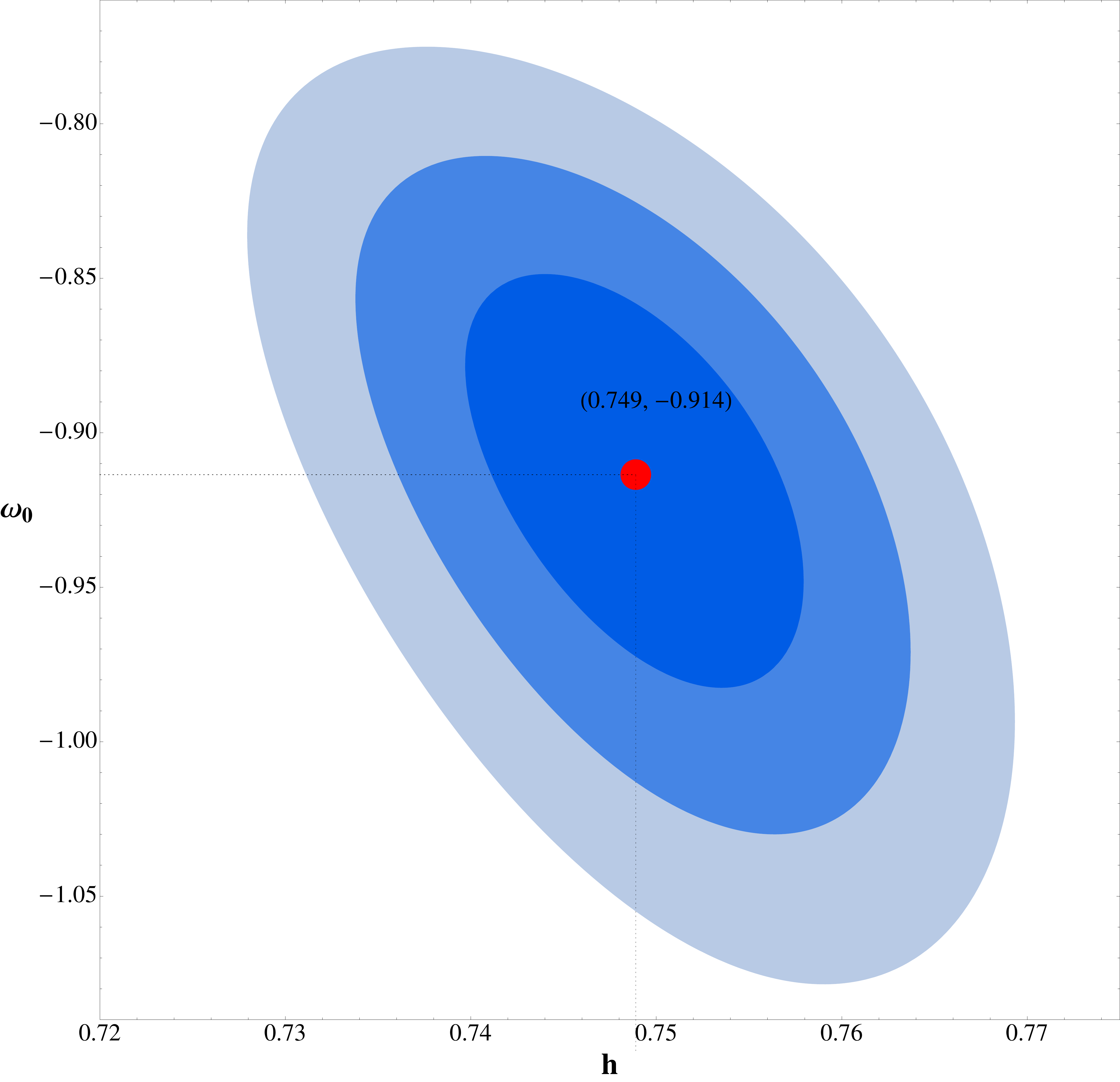}
\end{tabular}
\vspace{-0.5cm}
\caption{$\omega_0$, $h$, and $b$ constraints for $\omega$CDM model when $\Omega_{\m0} = 0.32$. a) The contours for $b$-$\omega_0$ constraint. The contour levels represent the 1-$\sigma$, 2-$\sigma$, and 3-$\sigma$ constraints. b) The constraint contours for $b$ and $h$. c) The contours for $h$-$\omega_0$ constraint.} \label{fig-3}
\vspace{1cm}
\end{figure*}
	\end{itemize}

\subsection{CPL  : $\omega_0 + \omega_a (1-a)$}
\label{subsec:CPL}
In this subsection, we investigate the so-called CPL model. The best fit values of the lens model parameters are almost the same for all models as shown in other subsections. We consider the different values of $\Omega_{\m0}$ models for the different choices of $\omega_0$. Unfortunately, we are not able to obtain the converge values of cosmological parameters when we also allow $\omega_0$ to vary. Thus, we investigate cosmological parameters for the given values of $\omega_0 = 0.95, -1.0$, and $-1.05$ with fixing $h = 0.74$. For the lower values of $h$, these models suffer from converging of the chi-square analysis.
	\begin{itemize} 
		\item \textbf{$\Omega_{\m0} = 0.27$ :} First, we investigate $\omega_0 = -0.95$. $\left( \gamma_0 \,, \gamma_{z} \,, \gamma_{s} \,, \beta \,, \delta \right) = \left( 1.198 \,, -0.560 \,, 0.813 \,, 2.354 \,, -0.023 \right)$. Within 1-$\sigma$ CL, we obtain $(b \,, \omega_a) = (-0.037 \pm 0.113 \,, -1.115 \pm 0.301)$ with $\chi^2 = 326.3$. The best-fit values of lens model parameters are almost the same as each other for the different values of $\omega_0$ in this model. The 1-$\sigma$ CL values of $( b \,, \omega_a) = (-0.053 \pm 0.103 \,, -0.992 \pm 0.247)$ with $\chi^2 = 350.2$ for $\omega_0 = -1.0$. Also, $b  = -0.057 \pm 0.090 $ with $\chi^2 = 338.4$ for $(\omega_0 \,, \omega_a ) = (-1.5 \,, -0.88)$. The analysis does not converge for the general $\omega_a$ case in the last model and thus we fix its value in this model. For these models, the data show no time variation of the speed of light because of the null results of $b$.
		\item \textbf{$\Omega_{\m0} = 0.32$ :}  $\left( \gamma_0 \,, \gamma_{z} \,, \gamma_{s} \,, \beta \,, \delta \right) = \left( 1.185 \,, -0.582 \,, 0.820 \,, 2.363 \,, -0.019 \right)$ for $\omega_0 = -0.95$. The 1-$\sigma$ CL best fit values of this model become $(b \,, \omega_a) = (-0.228 \pm 0.104 \,, -1.042 \pm 0.251)$ with $\chi^2 = 391.2$ for this model. The best fit values of lens model parameters are quite similar for the different values of $\omega_0$ in this model as shown in Table.~\ref{tab:omegaw0}. $(b \,, \omega_a \,, \chi^2) = (-0.250 \pm 0.103 \,, -0.922 \pm 0.247 \,, 400.5)$ and $ (-0.326 \pm 0.101 \,, -0.804 \pm 0.241 \,, 430.0)$ for $\omega_{0} = -1.00$ and $-1.05$, respectively. We show the 1-$\sigma$ confidence evolution plots for $\tc/\tc_0$ and $\tG/\tG_0$ in Fig.~\ref{fig-4}. In the left panel, the evolution of $\tc/\tc_0$ within 1-$\sigma$ CL is shown. The dashed line corresponds to the best-fit value of parameters. It can deviate from the present value of about 9\% at $z=3$. The cosmological evolution of $\tG/\tG_0$ is shown in the right panel of this figure. The value of the gravitational constant at $z= 3$ can be 40 \% larger than that of the present value in this model. This result is opposite to the results obtained from using the growth rate \cite{Lee:2012nh,Nesseris:2017vor,Kazantzidis:2018jtb} and from SNe data \cite{Lee:2021ona}.
\begin{figure*}
\centering
\vspace{1cm}
\begin{tabular}{cc}
\includegraphics[width=0.48\linewidth]{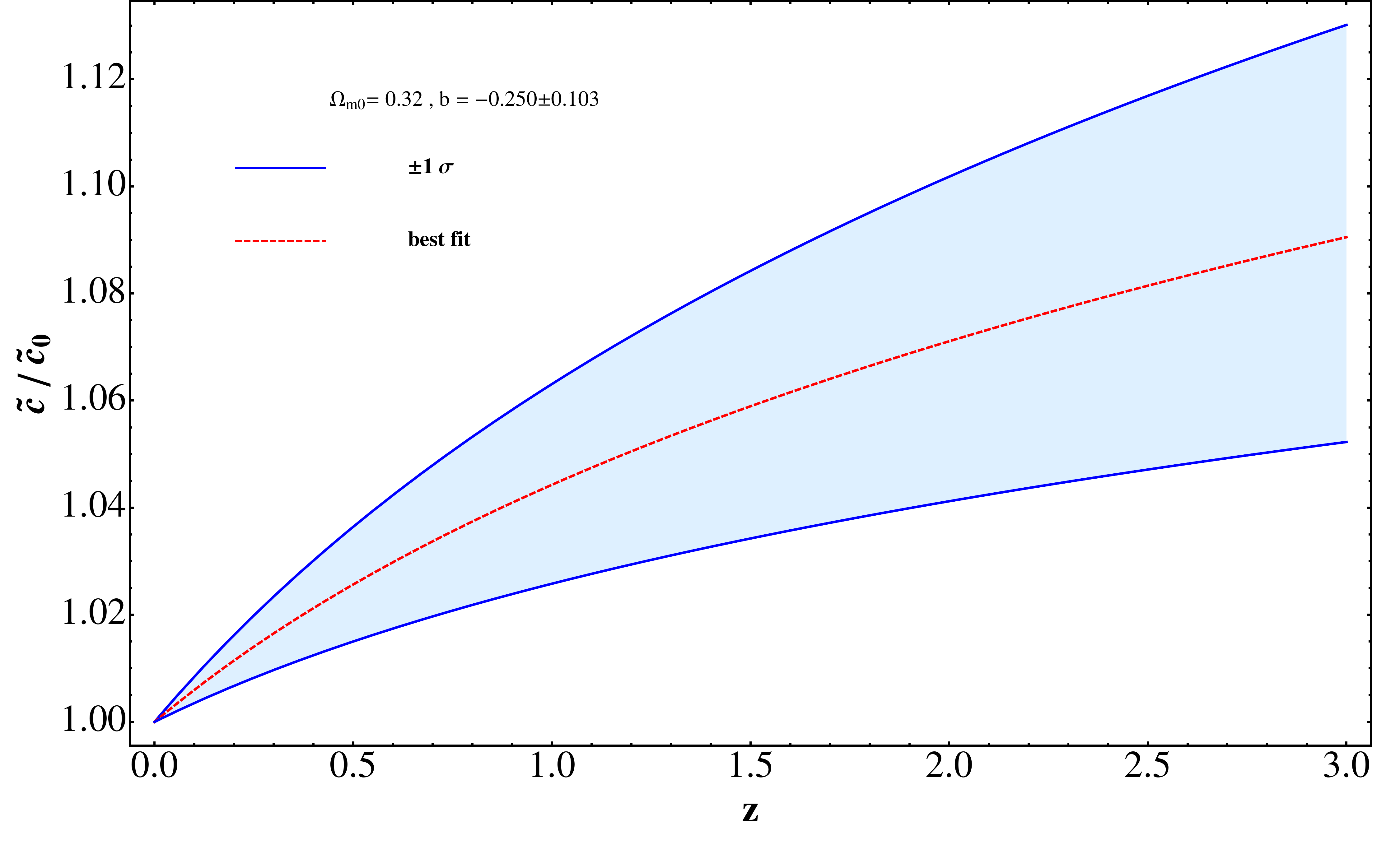} &
\includegraphics[width=0.48\linewidth]{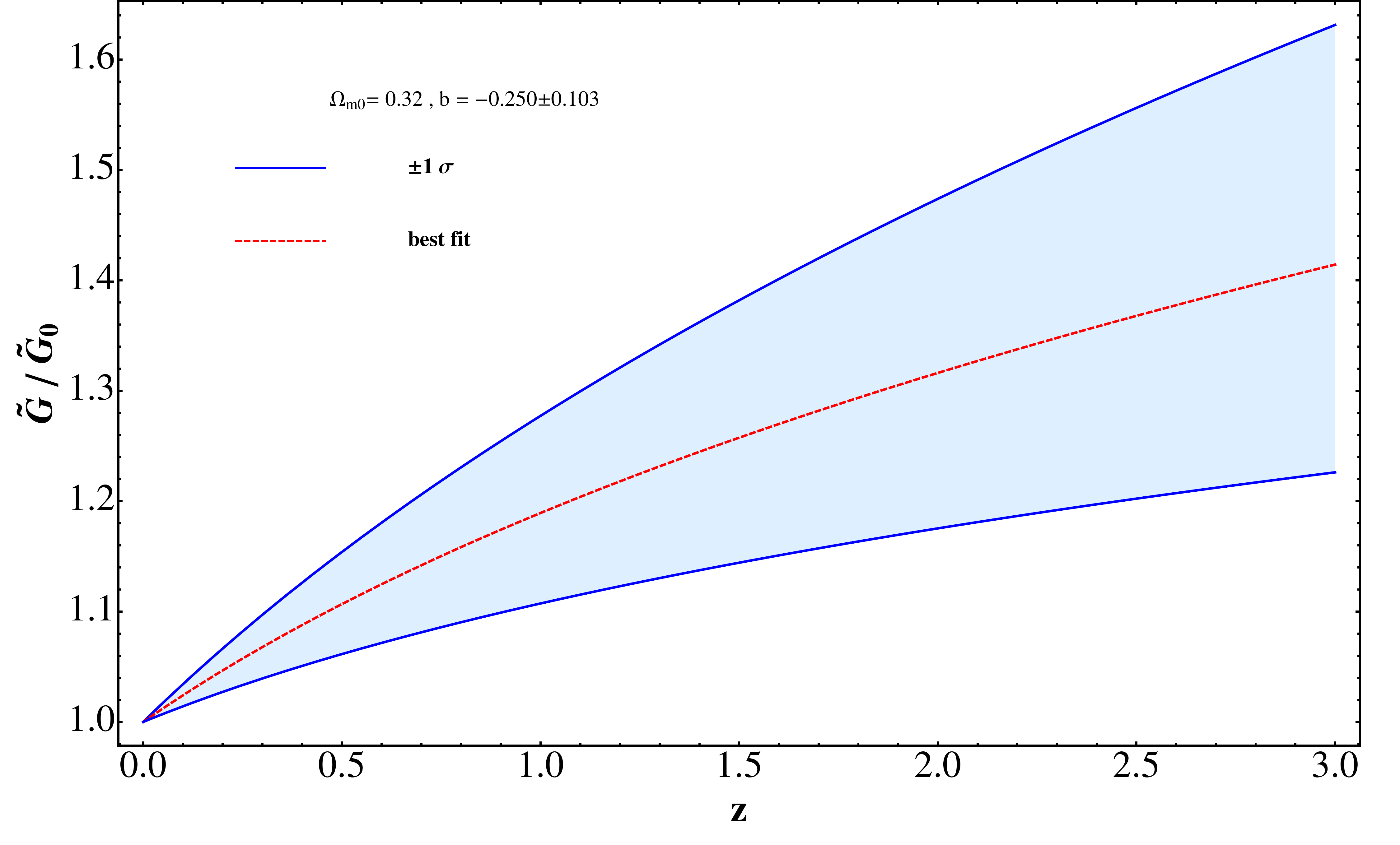} 
\end{tabular}
\vspace{-0.5cm}
\caption{The time variations of $\tc$ and $\tG$ of CPL model with $\omega_0 = -1.0$ as a function of $z$ when $\Omega_{\m0} = 0.32$. a) The cosmological evolution of the ratio of $\tc$ to its present value $\tc_0$ within 1-$\sigma$ CL. a) The cosmological evolution of the value of $\tG/\tG_0$ for 1-$\sigma$ CL.} \label{fig-4}
\vspace{1cm}
\end{figure*}
	\end{itemize}

\begin{table}[h!]
 	\begin{center}
		\caption{Best fit values and their 1-$\sigma$ errors of cosmological parameters for various DE models of meVSL. }
		\label{tab:omegaw0}
		\begin{tabular}{|c|c|c|c|c|c|c|c|c|c|c|} 
			\hline
			 $\omega_0$ & $\omega_a$ & $\Omega_{\m0}$ & $\gamma_0$ & $\gamma_{z} $ & $\gamma_{s}$  &  $\delta$ & $\beta$ & $h$& $b$  & $\chi^2$ \\ \hline
			  \multirow{4}{*}{-1} & \multirow{4}{*}{0} &\multirow{2}{*}{0.27}  & $1.201$ & $-0.568$ &$0.815$ &$2.349$ & $-0.022$ & $0.717$ & 0 & 344.8 \\  
			  & &$$  & $1.182$ & $-0.575$ &$0.818$ &$2.370$ & $-0.019$ & $0.738$ & $0.032 \pm 0.092$ & 359.6 \\ \cline{3-11}
			  & &\multirow{2}{*}{0.32}   & $1.184$ & $-0.589$ &$0.822 $ &$2.362$ & $-0.018$ & $0.721$ & 0 & 431.1 \\ 
			  & &  & $1.183$ & $-0.588$ &$0.822 $ &$2.363$ & $-0.018$ & $0.731$ & $-0.339 \pm 0.091$ & 421.4 \\ \hline 
			  $-0.953 \pm 0.044$ & \multirow{4}{*}{0} &\multirow{2}{*}{0.27}  & $1.197$ & $-0.570$ &$0.816$ &$2.352$ & $-0.021$ & $0.717$ & 0 & 345.8 \\  
			  $-1.006 \pm 0.062$ & &$$  & $1.198$ & $-0.574$ &$0.817$ &$2.367$ & $-0.006$ & $0.710$ & $0.127 \pm 0.120$ & 357.0 \\ \cline{3-11}
			  $-0.992 \pm 0.042$ & &\multirow{2}{*}{0.32}   & $1.181$ & $-0.591$ &$0.822 $ &$2.364$ & $-0.017$ & $0.721$ & 0 & 431.1 \\ 
			 $-0.966 \pm 0.052$ & &  & $1.205$ & $-0.590$ &$0.821 $ &$2.359$ & $-0.017$ & $0.700$ & $-0.246 \pm 0.108$ & 418.1 \\ \hline 
			    $-0.944 \pm 0.054$ & \multirow{2}{*}{0} & 0.27 &\multirow{2}{*}{1.18} & \multirow{2}{*}{-0.49} & \multirow{2}{*}{0.81} &\multirow{2}{*}{2.48} & \multirow{2}{*}{0}   & $0.743 \pm 0.009$ & $0.331 \pm  0.171$ & 557.4 \\ 
			    $-0.914 \pm 0.046$&  &0.32  &  &  & & &  & $0.749 \pm 0.009$ & $-0.214 \pm 0.166$ & 649.7 \\ \hline
			   $-0.95$ & $-1.115 \pm 0.301$ &\multirow{3}{*}{0.27}  & $1.198$ & $-0.560$ &$0.813$ &$2.354$ & $-0.023$ & \multirow{3}{*}{0.74}  & $-0.037 \pm 0.113$ & 326.3 \\  
			  $-1.00 $ & $-0.992 \pm 0.247$&$ $  & $1.199$ & $-0.560$ &$0.814$ &$2.354$ & $-0.023$ &  & $-0.053 \pm 0.103$ & 350.2 \\ 
			   $-1.05 $ & -0.88 &$$  & $1.182$ & $-0.563$ &$0.814$ &$2.354$ & $-0.022$ &  & $-0.057 \pm 0.090$ & 338.4 \\ \hline
			   $-0.95$ & $-1.042 \pm 0.251$ &\multirow{3}{*}{0.32}  & $1.185$ & $-0.582$ &$0.820$ &$2.363$ & $-0.019$ & \multirow{3}{*}{0.74}  & $-0.228 \pm 0.104$ & 391.2 \\  
			  $-1.00 $ & $-0.922 \pm 0.247$&  & $1.186$ & $-0.582$ &$0.820$ &$2.363$ & $-0.019$ &  & $-0.250 \pm 0.103$ & 400.5 \\ 
			   $-1.05 $ & $-0.804 \pm 0.241$ &$$  & $1.191$ & $-0.581$ &$0.820$ &$2.358$ & $-0.021$ &  & $-0.326 \pm 0.101$ & 430.0 \\ \hline
		\end{tabular}
	\end{center}
 \end{table}

\section{Conclusions}
\label{sec:Con}

We have analyzed a galaxy-scale strong gravitational lensing sample including 161 systems with stellar velocity dispersion measurements for the meVSL model. We adopt the lens model parameters with the slope of the total mass density profile $\gamma$ depending on the lens redshift. We obtain the best fit values of lens model parameters including the slope of the luminosity density, $\delta$ and that of the orbit anisotropy parameter $\beta$. All the parameters strongly depend on the value of the present values of the matter density contrast $\Omega_{\,m0}$ and thus we put prior on that value from WMAP and Planck separately.  

We investigate dark energy models for $\Lambda$, the constant $\omega$, and the CPL. When we adopt the value of $\Omega_{\m0}$ from the WMAP, strong gravitational lensing data provides the positive values of meVSL parameter, $b$ for $\Lambda$ and constant $\omega$ models. These indicate both the slower speed of light and the weaker gravitational force in the past. The result for the CPL model provides the negative values of $b$ indicating the faster speed of light and the stronger gravitational force in the past. However, the results for the $\Lambda$ and the CPL are null within 1-$\sigma$ CL. 

If we put the prior value of $\Omega_{\m0}$ from the Planck, then we obtain the negative values of meVSL parameter, $b$ for $\Lambda$, constant $\omega$ models, and CPL models. These mean both the faster speed of light and the stronger gravitational force in the past. These results are satisfied for all models within 1-$\sigma$ CL. However, these results are controversial with those obtained from the growth rate or supernovae data. 

This method includes the fitting of the lens model and this might provide the systematics to the whole analysis. Also, the statistical power of strong gravitational lensing is still not enough to remove prior to the value of $\Omega_{\m0}$ in the analysis. This is important because we obtain the opposite results for the value of $b$ for the different values of $\Omega_{\m0}$. Thus, we might need more data on the strong gravitational lensing to put more reliable constraints on the meVSL model.

\section*{Acknowledgments}
SL is supported by Basic Science Research Program through the National Research Foundation of Korea (NRF) funded by the Ministry of Science, ICT, and Future Planning (Grant No. NRF-2017R1A2B4011168).


\end{document}